\documentclass[journal,single-space,two column,twoside,10pt]{IEEEtran}
\usepackage{verbatim}
\usepackage{amsmath}
\usepackage{amsthm}
\usepackage{amssymb}
\usepackage{graphicx}
\usepackage{subfigure}
\usepackage{mathrsfs}
\usepackage{algorithm}
\usepackage{algorithmic}
\usepackage{color}
\usepackage{enumerate}
\usepackage{setspace}
\usepackage{balance}

\newtheorem{theorem}{Theorem}
\newtheorem{lemma}{Lemma}
\newtheorem{corollary}{Corollary}
\newtheorem{proposition}{Proposition}
\newtheorem{definition}{Definition}

\ifCLASSINFOpdf
\else
\fi
\usepackage{url}

\IEEEoverridecommandlockouts

\begin{document}

\title{GreenDCN: a General Framework for Achieving Energy Efficiency in Data Center Networks}

\author{
Lin~Wang, Fa~Zhang, Jordi~Arjona~Aroca, Athanasios~V.~Vasilakos,~\IEEEmembership{Senior Member,~IEEE}\\ 
Kai~Zheng,~\IEEEmembership{Senior Member,~IEEE}, Chenying~Hou, Dan~Li, and Zhiyong~Liu 
\thanks{
Manuscript received January 15, 2013; revised May 28, 2013.

L. Wang and C.-Y. Hou are with the Center for Advanced Computing Research, Institute of Computing Technology, Chinese Academy of Sciences and the University of Chinese Academy of Sciences, Beijing, China.

F. Zhang is with the Center for Advanced Computing Research, Institute of Computing Technology, Chinese Academy of Sciences, Beijing, China.

J. Arjona Aroca is with Institute IMDEA Networks and University Carlos III of Madrid, Madrid, Spain.

A. V. Vasilakos is with the University of Western Macedonia, Greece.

K. Zheng is with the IBM China Research Lab, Beijing, China.

D. Li is with Tsinghua University, Beijing, China.

Z.-Y. Liu is with State Key Laboratory for Computer Architecture, Institute of Computing Technology, Chinese Academy of Sciences, Beijing, China.  Z.-Y. Liu and F. Zhang are the corresponding authors. Email: \{zyliu, zhangfa\}@ict.ac.cn.
}}

\markboth{IEEE Journal on Selected Areas in Communications, VOL. ,NO. , January 2014}{WANG \lowercase{\textit{et al.}}: GreenDCN: a General Framework for Achieving Energy Efficiency in Data Center Networks}

\maketitle

\begin{abstract}

The popularization of cloud computing has raised concerns over the energy consumption that takes place in data centers. In addition to the energy consumed by servers, the energy consumed by large numbers of network devices emerges as a significant problem. Existing work on energy-efficient data center networking primarily focuses on traffic engineering, which is usually adapted from traditional networks. We propose a new framework to embrace the new opportunities brought by combining some special features of data centers with traffic engineering. Based on this framework, we characterize the problem of achieving energy efficiency with a time-aware model, and we prove its NP-hardness with a solution that has two steps. First, we solve the problem of assigning virtual machines (VM) to servers to reduce the amount of traffic and to generate favorable conditions for traffic engineering. The solution reached for this problem is based on three essential principles that we propose. Second, we reduce the number of active switches and balance traffic flows, depending on the relation between power consumption and routing, to achieve energy conservation. Experimental results confirm that, by using this framework, we can achieve up to $50\%$ energy savings. We also provide a comprehensive discussion on the scalability and practicability of the framework.

\end{abstract}


\section{Introduction}
\label{sec:intro}

\PARstart{D}{ata} centers are integrated facilities that house computer systems for cloud computing and have been widely deployed in large companies, such as Google, Yahoo! or Amazon. The energy consumption of data centers has become an essential problem. It is shown in \cite{Koomey-2011} that the electricity used in global data centers in 2010 likely accounted for between $1.1\%$ and $1.5\%$ of the total electricity use and is still increasing. However, while energy savings techniques for servers have evolved, the energy consumption of the enormous number of network devices that are used to interconnect the servers has emerged as a substantial issue. Abts et al. \cite{Abts_Marty-2010} showed that, in a typical data center from Google, the network power is approximately $20\%$  of the total power when the servers are utilized at $100\%$, but it increases to $50\%$ when the utilization of servers decreases to $15\%$, which is quite typical in production data centers. Therefore, improving the energy efficiency of the network also becomes a primary concern.

There is a large body of work in the field of energy efficiency in Data Center Networks (DCNs). While some energy-efficient topologies have been proposed (\cite{Abts_Marty-2010}, \cite{Huang_Jia-2011}), most of the studies are focused on traffic engineering and attempt to consolidate flows onto a subset of links and switch off unnecessary network elements (\cite{Wang_Yao-2012}, \cite{Heller_Seetharaman-2010}, \cite{Shang_Li-2010}, \cite{Zhang_Ansari-2012}). These solutions are usually based on characterizing the traffic pattern by prediction, which is usually not feasible or is not precise enough because the traffic patterns vary significantly depending on the applications.

We believe that, in order to improve the energy efficiency in DCNs, the unique features of data centers should be explored. More specifically, the following features are relevant: \\
a) \textit{Regularity of the topology:} compared to traditional networks, DCNs use new topologies, such as Fat-Tree \cite{Al-Fares_Loukissas-2008}, BCube \cite{Guo_Lu-2009} and DCell \cite{Guo_Wu-2008}, which are more regular and symmetric. As a result, it is possible to have better knowledge about the physical network.\\
b) \textit{VM assignment:} because of virtualization, we can determine the endpoints of the traffic flows, which will have a remarkable influence on the network traffic and will, consequently, condition the traffic engineering.\\
c) \textit{Application characteristics:}  most applications in cloud data centers are run under the MapReduce paradigm \cite{Dean_Chemawat-2008}, which can cause recognizable communication patterns. Making use of these characteristics can help eliminate the need for traffic prediction and obtain better traffic engineering results.

\begin{figure}
\centering
	\includegraphics[scale=0.5]{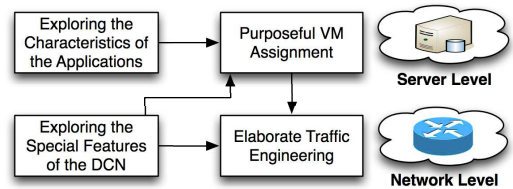}
   \caption{\label{fig:framework}A general framework for improving the energy efficiency in DCNs.}
	\vspace{-0.6cm}
\end{figure}

To take full advantage of these new opportunities, we propose a new general framework (as illustrated in Figure~\ref{fig:framework}) for achieving energy efficiency in DCNs, where the specific information on both the applications and the network will be deeply explored and coherently utilized. 
We will carefully design the VM assignment based on a comprehensive understanding of the applications' characteristics and combine them with the aforementioned network features (e.g., topology, end-to-end connectivity). This purposeful VM assignment will provide us with favorable traffic conditions on the DCN and, thus, gain some energy savings in advance before performing traffic engineering on the network. Then, we will explore specific traffic engineering solutions according to the specific traffic patterns and network features.

The main contributions of this paper are highlighted as follows. First, we provide a new general framework for energy minimization in DCNs. We also conduct exhaustive analysis on how to proceed with this framework and identify new issues and challenges. Second, we model the energy-saving problem in DCNs by using this new framework and analyze its complexity. Third, we provide in-depth analysis on both VM assignment and network routing with respect to energy conservation, showing that there is much room for improving the energy efficiency by making use of some unique features of data centers. Fourth, based on the analytical results, we provide efficient algorithms to solve the problem. We also conduct comprehensive experiments to evaluate the efficiency of our method. 

The remainder of this paper is organized as follows. In Section~\ref{sec:framework}, we describe the general framework and discuss how it can be deployed. In addition, we list some newly arising issues. In Section~\ref{sec:model}, we present a time-aware model to describe the energy-saving problem in DCNs based on the new framework and analyze its complexity. We explore VM assignment principles for energy saving and provide a traffic-aware energy-efficient VM assignment algorithm in Section~\ref{sec:assign}. The routing optimization is addressed in Section~\ref{sec:route}, where we present detailed theoretical analysis and provide a two-phase energy-efficient routing algorithm. Section~\ref{sec:evaluation} provides the experimental results, and Section~\ref{sec:discussion} presents some extended discussion on the practicality of our algorithms. In Section~\ref{sec:related}, we summarize related studies, and in Section~\ref{sec:conclusion}, we draw final conclusions. All of the proofs for the lemmas and theorems in this paper are given in the Appendix.

\section{The General Framework}
\label{sec:framework}

Although we consider the problem of achieving energy efficiency in DCNs, this framework can be generalized for most performance optimization problems in DCNs. In this section, we discuss in general how to conduct optimization work by using this framework, and we identify some new challenges. The structure of this new framework is illustrated in Figure~\ref{fig:framework}.

\textbf{Applications.} As an important paradigm for large-scale data processing, MapReduce \cite{Dean_Chemawat-2008} has been widely applied in modern cloud data centers.
Most cloud applications have been ported to MapReduce. For this reason, we focus on typical MapReduce jobs. A typical MapReduce job comprises three main phases: Map, Shuffle and Reduce. The network is intensively used only in the Shuffle phase to exchange intermediate results between servers. As a result, MapReduce-type applications usually show regular communication patterns. Xie et al. \cite{Xie_Ding-2012} profiled the network patterns of several typical MapReduce jobs, including Sort, Word Count, Hive Join, and Hive Aggregation, which represent an important class of applications that reside in data centers. They observed that all of these jobs generate substantial traffic during only $30\%$-$60\%$ of the entire execution. The traffic patterns of these jobs can mainly be classified into three categories: single peak, repeated fixed-width peaks and varying height and width peaks. Having these patterns in mind, the network traffic can be scheduled in advance, which will condition the traffic engineering results.

The characteristics of applications can be obtained by profiling runs of jobs. The detailed profiling method is beyond the scope of this paper, but one possible realization can be found in \cite{Xie_Ding-2012}. The profiling process can bring ineluctable profiling overhead, but it can be drastically reduced if the same types of jobs with the same input size are run repeatedly. We observe that such a scenario is quite common in cloud data centers for iterative data processing such as PageRank \cite{Page_Brin-1999}, where much of the data remains unchanged from iteration to iteration, and also in many production environments (e.g., \cite{Agarwal_Kandula-2012}), the same job must be repeated many times with almost identical data.

\textbf{Data center networks.} To provide reliability and sufficient bisection bandwidth, many researchers have proposed alternatives to the traditional 2N tree topology \cite{Cisco}. By providing richer connectivity, topologies such as Fat-Tree (\cite{Al-Fares_Loukissas-2008, Niranjan_Pamboris-2009}), BCube \cite{Guo_Lu-2009}, DCell \cite{Guo_Wu-2008} and VL2 \cite{Greenberg_Hamilton-2009} can handle failures more gracefully. Among them, Fat-Tree was proposed to use commodity switches in data centers, which can support any communication pattern with full bisection bandwidth.

Furthermore, the DCN provides another special benefit: regularity of the topology. Most of the topologies that are being used in DCNs follow a multi-tier tree architecture. The scalability of such topologies is always achieved by scaling up each individual switch, i.e., by increasing the fan-out of single switches rather than scaling out the topology itself. Because such topologies in different scales always possess almost the same properties, the optimization efforts that we make for small-scale networks can be easily adapted to large-scale networks with very slight changes. This arrangement enables us to make use of the unique features of well-structured topologies to improve network performance by gaining insights from small-scale networks.

\textbf{VM assignment.} To improve the flexibility and overall hardware-resource utilization, virtualization has become an indispensable technique in the design and operation of modern data centers. 
Acting as a bridge, VM assignment provides the possibility of combining application characteristics and traffic engineering. With the goal of improving the network performance, an efficient VM assignment can be achieved by integrating the characteristics of the running applications and the special features of the network topology. For example, knowing the traffic patterns of applications, we can schedule jobs such that their communication-intensive periods are staggered, or jobs with similar communication patterns are separated into different areas of data centers. As a consequence, the load on the network will be more balanced, and the network utilization will be accordingly improved. By assigning VMs in an appropriate way, we will be able to obtain better initial conditions for the subsequent traffic engineering.

\textbf{Traffic engineering.} As a conventional approach for the optimization of network performance, traffic engineering has also been extensively investigated in DCNs. Most of the traffic engineering solutions being used in current data centers are simply adapted from traditional networks. In a traditional operational network, traffic engineering is usually conducted by traffic measurement, characterization, modeling and control. However, with the specific features that characterize DCNs, traffic engineering could be quite different from the conventional cases. Using the information on traffic patterns provided by VM assignment, a better understanding of the traffic can be achieved and, consequently, traffic measurement and characterization can be eliminated, which could lead to more precise traffic engineering results. At the same time, we can also take advantage of the unique features of the DCN topology and design elaborate traffic engineering solutions more specifically.

Under this new framework, there are some newly arising issues and challenges that could require future research efforts: 
a) The applications running in current data centers show regular communication patterns and can be obtained by profiling. However, the profiling method will directly condition the accuracy of this information. As a result, effective and efficient profiling methods are highly desired.
b) Different metrics for network performance could prefer different favorable traffic conditions, which are conditioned by VM assignment. Thus, understanding favorable traffic conditions and designing efficient VM assignment algorithms to generate them will be crucial in this framework. 
c) Universal traffic engineering solutions might not be efficient enough for current DCNs. To obtain better results, specific traffic engineering methods for each specific data center must be explored by making use of both the topology features and the traffic patterns that are known in advance.

\section{Modeling the Energy-Saving Problem}
\label{sec:model}

We present a temporal model for the energy-saving problem and analyze its complexity in this section.

\vspace{-0.3cm}

\subsection{Data Center and Data Center Network}
\label{sec:DCandDCN}

We consider a data center to be a centralized system in which a set of servers is connected by a well-designed network. Assume that there is a set of servers that are represented by $\mathcal{S}$. To achieve better utilization of the hardware resources, the jobs are processed by VMs that are hosted by servers. All of the servers are connected by a network $\mathcal{G}=(\mathcal{V},\mathcal{E})$, where $\mathcal{V}$ is the set of network devices\footnote{Because the network devices are mainly switches, from now on, we will use the term \emph{switches} instead of \emph{network devices}.}, and $\mathcal{E}$ is the set of links. In this work, we focus on switch-centric physical network topologies and use the most representative one, Fat-Tree, to conduct our work. For each switch $v \in \mathcal{V}$, the total traffic load that it carries can be expressed by
$x_{v} = \frac{1}{2} \sum_{\{e \in \mathcal{E} : e \text{ is incident to }v\}} y_e,$
where $y_e$ is the total traffic carried by link $e$. A factor of $1/2$ is necessary to eliminate the double counting of each flow because each flow that arrives at a node must also depart.

For single network elements, energy-saving strategies have been widely explored. Among them, \emph{speed scaling} (\cite{Yao_Demers-1995, Bansal_Kimbrel-2007, Gunaratne_Christensen-2008, Andrews_Fernandez-2010}) and \emph{power down} (\cite{Nedevschi_Popa-2008, Parnaby_Zimmerman-2008}) are two representative techniques. 
In this paper, we use both strategies in an integrated way. More precisely, we characterize the power consumption of a switch $v \in \mathcal{V}$ by an energy curve $f_v(x_v)$, which indicates how $v$ consumes power as a function of its transmission speed $x_v$. Usually, function $f_v(x_v)$ can be formalized as
\begin{equation}
	\label{eq:cost_function}
	f_v(x_v)=\left\{
	\begin{aligned}
		& 0 & \mbox{for } x_v = 0 \\
		& \sigma_v + \mu_v x_v^{\alpha} & \mbox{for } x_v > 0
	\end{aligned}
	\right.,
\end{equation}
where $\sigma_v$ represents the fixed amount of power needed to keep a switch active, while $\mu_v$ and $\alpha$ are parameters that are associated with the switches. In this way, if a switch carries no load, then it can be shut down and incurs no cost. Otherwise, an initial cost is paid at the beginning, and then the cost increases as the assigned load increases. We assume that the power consumption of a switch grows superadditively with its load, with $\alpha$ usually being larger than $1$ \cite{Andrews_Fernandez-2010}. Due to the homogeneity in DCNs, it is convenient to assume that there is a uniform cost function $f(\cdot)$ for all of the switches. The total cost of a network is defined as the total power consumption of all of its switches, which is given by $\sum_{v \in \mathcal{V}} f(x_v)$.

\vspace{-0.2cm}

\subsection{Applications}

As we discussed before, the applications can be mainly classified into three categories according to their communication patterns. We choose the most general communication pattern, which is varying height and width peaks, to build our model. This communication pattern assumes that there can be multiple communication-intensive periods during the execution of a job and that the lengths of these periods, as well as the traffic generated in different periods by this job, can be different.

Assume that we are given a set $\mathcal{J}$ of jobs that have to be processed simultaneously during the time period of interest $[t_1,t_{r}]$.
We choose timeslots such that during each timeslot, the traffic is relatively stable.
Each job $j \in \mathcal{J}$ is composed of $n_j$ tasks that will be processed on a pre-specified VM $m$ from the set of VMs $\mathcal{M}$. For each job $j$, there is a traffic matrix for its $n_j$ associated VMs, denoted by $\mathbf{T}_j(t)$, where $t \in [t_1,t_r]$ is a timeslot.

We assume that the communication of a job is concentrated in certain timeslots.
We call each continuous communication-intensive period a \emph{transfer}. Formally, for each job $j$, we define
\begin{equation}
\mathcal{T}_j = \left\{ (t_{ji}^{start}, t_{ji}^{end}, \mathbf{B}_{ji})~|~i \in [1,L_j] \right\}
\end{equation}
that contains $L_j$ transfers that are given by $3$-tuples. In each $3$-tuple,
$t_{ji}^{start}$ and $t_{ji}^{end}$ represent the start and end time of the $i$-th transfer, respectively, while
$\mathbf{B}_{ji}$ denotes the traffic matrix of the VMs
that are present in this transfer, i.e., $\mathbf{T}_j(t) = \mathbf{B}_{ji}$ if timeslot $t \in [t_{ji}^{start}, t_{ji}^{end}]$. We assume that, for any timeslot $t \not \in [t_{ji}^{start}, t_{ji}^{end}]$, there is only background traffic, i.e., $\mathbf{T}_j(t)=\epsilon\rightarrow0$, and it has only a small influence on the network.

\subsection{Problem Description}

Next, we describe the energy-saving problem in DCNs and provide a time-aware network energy optimization model to redefine this problem. We assume that the VMs
will not be migrated once they have been assigned because in cloud data centers, jobs are usually very small \cite{Xie_Ding-2012}. For example, the average completion time of a MapReduce job at Google was $395$ seconds during September 2007 \cite{Dean_Chemawat-2008}. 

The total energy consumed by all of the switches for processing all of the jobs can then be represented by
\begin{equation}
E = \sum_{t = t_1}^{t_{r}} \left( \sum_{v \in \mathcal{V}} f(x_v(t))\right) \label{eq:total_energy},
\end{equation}
where $x_v(t)$ is the load of switch $v$ in timeslot $t$. Our goal is to assign all of the VMs to servers such that when we choose appropriate routing paths for the flows between each pair of VMs, the total cost $E$ is minimized.

The optimization procedure can be divided into two closely related stages: VM assignment and traffic engineering. Given an assignment of VMs, the total cost can be minimized by applying traffic engineering on the network, which solves the energy-efficient routing problem. We first assume that an algorithm $\mathbb{A}$ has been given to solve this routing problem. Then, the VM assignment problem can be modeled by the following integer program:
\begin{equation}
	\begin{array}{lll}
		& (IP_1) \;\; \min \;\; \sum_{t = t_1}^{t_{r}} \mathbb{A}(\mathcal{D}(t)) \nonumber \\
		\mbox{ subject to} \nonumber\\
		& \sum_m \Delta_{m,s} \cdot C_m \le C_s & \forall s  \nonumber \\
		& \sum_{s \in \mathcal{S}} \Delta_{m,s} = 1 & \forall m \nonumber \\
		& \Delta_{m,s} \in \{0,1\} & \forall m,s  \nonumber
	\end{array}
	\nonumber
\end{equation}
where $\Delta_{m,s}$ indicates whether VM $m$ is assigned to server $s$. Variable $C_m$ represents the abstract resources that are required by VM $m$, and $C_s$ is the total amount of resources in one server. The second constraint means that each VM must be assigned to only one server. $\mathcal{D}(t)$ is a set of traffic demands to be routed in timeslot $t$. Each demand in $\mathcal{D}(t)$ is described by a triple that is composed of a source, a destination and a flow amount. Once an assignment is given, $\mathcal{D}(t)$ can be obtained by the active transfer of jobs.

Next, we discuss the energy-efficient routing problem that algorithm $\mathbb{A}$ aims to solve. After obtaining the traffic demands $\mathcal{D}(t)$, this problem can be represented as follows: given a network $\mathcal{G}=(\mathcal{V},\mathcal{E})$ with a node cost function $f(\cdot)$ and a set of traffic demands $\mathcal{D}(t)$, the goal is to inseparably route every demand in $\mathcal{D}(t)$ such that the total cost of the network $\sum_{v \in \mathcal{V}} f(x_v)$ is minimized, where $x_v$ is the total load of node $v$. Formally, this process can be formulated with the following integer program.
\begin{equation}
	\begin{array}{lll}
		& (IP_2) \;\; \min \;\; \sum_{v \in \mathcal{V}} f(x_v) \nonumber\\
		\mbox{subject to} \nonumber \\
		& x_v = \frac{1}{2} \sum_{e \in \mathcal{E}: e \text{ is incident to } v}y_e & \forall v \nonumber\\
		& x_v \leq C & \forall v \\
		& y_e = \sum_{d \in \mathcal{D}(t)} |d| \cdot \Phi_{d,e} &  \forall e \nonumber\\
		& \Phi_{d,e} \in \{0,1\} &  \forall d,e \nonumber\\
		& \Phi_{d,e}: \mbox{ flow conservation}\nonumber
	\end{array}
	\nonumber
\end{equation}
where $\Phi_{d,e}$ is an indicator variable that shows whether the demand $d \in \mathcal{D}(t)$ goes through edge $e$. The $1/2$ factor avoids counting each flow twice, as stated in subsection \ref{sec:DCandDCN}. Flow conservation means that only a source (sink) node can generate (absorb) flows, while for the other nodes, the ingress traffic equals the egress traffic. Variable $y_e$ is the total load carried by link $e$, and $x_v$ is the total traffic going through node $v$, which will never exceed the switch capacity $C$.

\subsection{Complexity Analysis}

We now analyze the computational complexity of this problem. In fact, the NP hardness can be proved by a reduction from the general Quadratic Assignment Problem (QAP), which describes the following problem: there is a set of $n$ facilities and a set of $n$ locations. A distance and a weight are specified for each pair of locations and facilities, respectively. The problem is to assign all of the facilities to different locations with the goal of minimizing the sum of the distances multiplied by the corresponding weights. QAP was first studied by Koopmans and Beckmann \cite{Koopmans_Beckmann-1957} and is a well-known strong NP-hard problem. Moreover, achieving any constant approximation for the general QAP is also NP-hard. It is believed that even obtaining the optimal solution for a moderate scale QAP is impossible \cite{Meng_Pappas-2010}. Formally, we show the following:

\begin{theorem}
Finding the optimality of the energy-saving problem in DCNs is NP-hard.
\end{theorem}

\section{Exploring Energy-Efficient VM Assignments}
\label{sec:assign}

In this section, we seek energy-efficient VM assignment strategies by exploiting some unique features of the usually well-structured topologies of DCNs. Combining this goal with the analysis of the characteristics of the applications, we provide three main principles to guide VM assignment. Based on these principles, we propose a traffic-aware energy-efficient VM assignment algorithm. \textcolor{black}{We first provide the following definitions:}

\begin{definition}
The \textbf{power rate} of a switch is defined as the power consumed by every unit of its load, i.e., $f(x)/x$ $(x > 0)$.
\end{definition}

\begin{proposition}
\label{prop:opt_load}
The total power consumption of a network is minimized when the number of active switches is optimum and their load is evenly balanced and as close to $R^* =  \left(\frac{\sigma}{\mu(\alpha-1)}\right)^{1/\alpha}$ as possible.
\end{proposition}

However, this proposition might not be directly applicable in reality. According to the statistics in \cite{Mahadevan_Sharma-2009}, the idle power consumption of a $48$-port edge LAN switch \textcolor{black}{usually} ranges from $76$ watts to $150$ watts, \textcolor{black}{increasing by approximately $40$ watts or more when running at full speed}. In \cite{Wang_Yao-2012}, the authors measured the power consumption of a production PRONTO $3240$ OpenFlow-enabled switch and obtained similar results. We also collected the power rating profiles of some commodity switches from vendors' websites; detailed information can be found in Table~\ref{tab:power}. We can see that the idle power usually occupies a large portion of the total power consumption, which means that the startup cost $\sigma$ in our model will be quite high. As a result, we will usually find that $R^* > C$. However, because the load in a switch cannot be larger than $C$, Proposition~\ref{prop:opt_load} might not apply. To consider this finding, we will assume in the remainder of this work that $R^* > C$, which in turn \textcolor{black}{assumes that} $\sigma > \mu(\alpha - 1) C^{\alpha}$.

\begin{table}[t]
\caption{\label{tab:power} Power rating profiles of some typical commodity switches (Unit: watts)}
\centering
\begin{tabular}{|c|c|c|}
	\hline
	\textbf{Product} & \textbf{Idle or Nominal} & \textbf{Max} \\
	\hline
	Cisco Nexus 3548 & 152 & 265 \\
	\hline
	Cisco Nexus 5548P & 390 & 600 \\
	\hline
	HP 5900AF-48XG & 200 & 260 \\
	\hline
	HP 5920AF-24XG & 343 & 366 \\
	\hline
	Juniper QFX 3600 & 255 & 345 \\
	\hline
\end{tabular}
\end{table}

\subsection{VM Assignment Principles for Saving Energy}
\label{sec:principles}

We now propose three principles for VM assignment that are intended to achieve a better energy efficiency in DCNs. We use a \emph{bottom-up} analysis approach, i.e., in a Fat-Tree, we focus on racks, on pods and finally on the whole data center.

\subsubsection{Minimizing energy at the rack level}

We first concentrate on determining the optimal number of Top-of-Rack (ToR) switches because ToR switches are different from other switches in the network. Once there is at least one active server in the rack, the corresponding ToR switch cannot be shut down because there might be some inter-rack traffic. ToR switches also carry intra-rack traffic, which will not be forwarded to other switches. As a result, the power consumption of the ToR switches will be largely conditioned by the VM assignment. The following theorem introduces how to assign VMs to racks.
\begin{theorem}
\label{theo:princ1}
(\textbf{Principle 1}) The optimal VM assignment compacts VMs into racks as tightly as possible to minimize the power consumption of the ToR switches.
\end{theorem}

\subsubsection{Minimizing energy at the aggregation level}

We now attempt to minimize the energy consumption at the aggregation level by choosing the optimal VM assignment and assuming that Theorem \ref{theo:princ1} is being applied; as a result, no ToR switches can be switched off again.
We assume a scenario in which there are a few jobs whose VMs are assigned to one pod and only one job is transferring at a certain timeslot. The next theorem follows:
\begin{theorem}
\label{theo:prin2}
(\textbf{Principle 2}) Distributing the VMs into $k$ racks results in less power consumption than compacting the VMs into a single rack, where $K$ is the number of racks in one pod and $4^{\frac{\alpha}{\alpha - 1}} \leq k \leq K$.
\end{theorem}

Theorem \ref{theo:prin2} implies that distributing the VMs among multiple racks will move some traffic from the ToR switches to the upper-layer network. Hence, because of the rich connectivity in the upper-layer network and the convexity property of energy consumption, a significant reduction in power consumption can be achieved; for example, when $\alpha = 2$, evenly distributing a job's VMs into $k=16$ or more racks will reduce the energy consumption compared to compacting the jobs into one rack. Because we are considering production data centers, $k>16$ in one pod is quite realistic as well as, in general, claiming that $K$ will not be smaller than $4^{\frac{\alpha}{\alpha - 1}}$. Note that if the inter-rack traffic is small, the energy saved will be even more significant.

Assuming, as above, that all VMs from the same job fit in the same rack is realistic. As was noted in \cite{Xie_Ding-2012}, most jobs in a large-scale data center can be fully assigned to a single rack and, in general, there will be few jobs that share a link at the same time. This last feature is, in fact, very important for us because, in our model, we will assign jobs that have complementary traffic patterns to the same pod. In this way, the interference between different jobs can be highly reduced.

\subsubsection{\textcolor{black}{Minimizing energy at the pod level}}

We now study how to assign VMs among different pods and whether it is better to assign all of a job's VMs to different pods or keep them together in one single pod. The next theorem provides the answer.

\begin{theorem}
(\textbf{Principle 3}) An optimal assignment will keep the VMs from the same job, if feasible, in the same pod.
\end{theorem}

\subsection{Energy-Efficient VM Assignment}

We devise an optimized energy-efficient VM assignment algorithm (optEEA) that was based on the three proposed principles. This algorithm will assign VMs with favorable traffic patterns for saving energy on the network by perfectly observing these principles. The algorithm takes a set of jobs (sets of VMs), its traffic patterns and a set of servers as input. Then, it returns job assignments (VM assignments) after going through the three steps listed in Section \ref{sec:principles}, which can also be seen in Algorithm~\ref{alg:vma}, lines $2$,$4$ and $6-7$.

\setlength{\textfloatsep}{10pt}
\begin{algorithm}[!t]
\caption{\label{alg:vma} \textbf{optEEA}}
\textbf{Input: } topology $\mathcal{G}=(\mathcal{V},\mathcal{E})$, servers $\mathcal{S}$ and jobs $\mathcal{J}$\\
\textbf{Output: } Assignments of VMs $\mathcal{M}$

\begin{algorithmic}[1]
\FOR{$j \in \mathcal{J}$}
	\STATE Transform VMs into super-VMs
\ENDFOR
\STATE Cluster jobs in $\mathcal{J}$ into groups $\mathcal{H}_i$ for $i \in [1,N^{pod}]$ and $\mathcal{H}_{N^{pod} + 1}$ 
\FOR{$1 \leq i \leq N^{pod}$}
	\STATE Partition the super-VMs for each job $j \in \mathcal{H}_j$ into $K$ parts using the min-$k$-cut algorithm
	\STATE Assign super-VMs to servers according to the partition
\ENDFOR
\STATE Assign the VMs of jobs in $\mathcal{H}_{N^{pod} + 1}$ into vacant servers in the first $N^{pod}$ pods flexibly.
\end{algorithmic}
\end{algorithm}

\textbf{First, transforming VMs into super-VMs.} Allowing each server to host multiple VMs would bring a high level of complexity to the subsequent steps. The transformation is conducted by following the proposition below.
\begin{proposition}
\label{prop:transformation}
Compacting the VMs that have a high level of communication traffic will reduce the network power consumption.
\end{proposition}

To complete this transformation, we define a referential traffic matrix $\mathbf{T}_j^{ref}$ for each job $j \in \mathcal{J}$, where
\begin{equation}
	\mathbf{T}_j^{ref}(m_1,m_2) = \sum_{t = t_1}^{t_r} \mathbf{T}_j(t)(m_1,m_2)
\end{equation}
for any $ m_1,m_2 \in [1,n_j]$. The referential matrix is used to indicate the total traffic generated from any VM to another VM for this job during the whole job lifetime.
For each job $j \in \mathcal{J}$, we shrink VMs to super-VMs by running the following process iteratively: 1) Choose the greatest value in matrix $\mathbf{T}_j^{ref}$. Assume that this value is located in the $m_1$-th row and the $m_2$-th column. 2) Combine the $m_1$-th VM with the $m_2$-th VM by removing the traffic between them and adding up their traffic with other VMs. 3) Choose the largest value in the $m_1$-th row and $m_2$-th row, and combine the corresponding VMs. We denote the VM that results after this shrinking process as a super-VM. We repeat this procedure until the resulting super-VM is large enough to exhaust the resources of a server. Then, we remove from the matrix all of the VMs that have been chosen and shrunk, and we find the next largest value to start a new iteration. With this transformation, all of the jobs will be represented by super-VMs, with each super-VM assigned to a single server.

\textbf{Second, clustering jobs into different pods.} We start by assuming that every job can be accommodated in a single pod. Nevertheless, if there are very large jobs that require more than one pod, we assign them in a greedy way, and then we consider assigning the remaining normal jobs. From Principles 1 and 3, we know that the number of pods that are used for accommodating all of the jobs must be minimized. In other words, it is not wise to separate the super-VMs for the same job into different pods if this job can be assigned into a single pod. Based on this consideration, we estimate the number of pods to be used by summing up the resources that are requested by all of the jobs. We denote the estimated number of pods as $N^{pod}$. Then, we partition the set of the jobs into those $N^{pod}$ pods by using a revised $k$-means clustering algorithm that takes the traffic patterns of the jobs into account. With the intuition that it is better to consolidate jobs that have strongly different traffic patterns into the same pod to improve the utilization of the network resources, the algorithm will compare the traffic patterns of the jobs and cluster them into different groups, where the difference in the communication patterns of the jobs in each group will be maximized.

To accomplish this goal, we first calculate a traffic pattern vector $\vec{\varphi}_j$ that has size $r$ for each job $j \in \mathcal{J}$. Each dimension of $\vec{\varphi}_j$ indicates the average traffic between any two VMs of job $j$ in each timeslot and is calculated as
\begin{equation}
\mathbf{T}_j^{avg}(t) =
 \frac{\sum_{m_1,m_2 \in [1, n_j]}\mathbf{T}_j(t)(m_1,m_2)}{n_j^2 / 2},
\end{equation}
if $t\in [t_j^s,t_j^t]$; otherwise, we set $\mathbf{T}_j^{avg}$ to $\epsilon$, where $\epsilon$ is infinitesimal. The traffic pattern vector now can be expressed as
\begin{equation}
\vec{\varphi}_j = \left(\mathbf{T}_j^{avg}(t_1), \mathbf{T}_j^{avg}(t_2),...,\mathbf{T}_j^{avg}(t_{r})\right).
\end{equation}
We then give the following definition:
\begin{definition}
Given two jobs $j_1, j_2 \in \mathcal{J}$ with traffic pattern vectors $\vec{\varphi}_{j_1}$ and $\vec{\varphi}_{j_2}$, respectively, the distance between the two jobs is defined as
\begin{equation}
dis(j_1,j_2) = dis(\vec{\varphi}_{j_1}, \vec{\varphi}_{j_2}) = \frac{1}{||\vec{\varphi}_{j_1} - \vec{\varphi}_{j_2}||_2}.
\end{equation}
\end{definition}
This definition of distance assumes that any two jobs that have similar traffic patterns will have a large distance between them. Having these distance vectors, the job clustering algorithm works as follows: 1) Choose $N^{pod}$ jobs and put them into sets $\mathcal{H}_i$ for $i \in [1,N^{pod}]$ with one job per set, using the traffic pattern vectors of those jobs as center vectors $\vec{\beta}_i$ of those sets. We adopt this initializing step from the refined $k$-means++ algorithm \cite{Arthur_Vassilvitskii-2007}. 2) For each of the remaining jobs $j$, find the nearest cluster $i$ with respect to the distance $dis(\vec{\varphi}_j, \vec{\beta}_i)$. If this job can be accommodated into this cluster without any resource violation, then put this job into set $\mathcal{H}_i$. Otherwise, choose the next job that has the largest distance and repeat this process until there is one cluster that can accommodate it. 3) Update the center vector of cluster $i$ by averaging all of the vectors of jobs in set $\mathcal{H}_i$,
\begin{equation}
	\vec{\beta}_i = \frac{\sum_{j \in \mathcal{H}_i} \vec{\varphi}_j}{|\mathcal{H}_i|} \;\;\;\;\;\;\;\;\forall i \in [1,N^{pod}].
\end{equation}
Repeat 2) and 3) until all of the jobs have been assigned. If there are some jobs that cannot find any cluster to accommodate them, then put them into an extra set $\mathcal{H}_{N^{pod} + 1}$. Finally, we choose $N^{pod}$ free pods and assign the jobs in each cluster to these pods.

\textbf{Third, assigning super-VMs to racks.} Inspired by Principle 2, we distribute the super-VMs of each job into multiple racks. The simplest way is to randomly partition these super-VMs into $K$ racks, where $K$ is the total number of racks in one pod. However, as we have stated before, it is better to allocate the VMs with the highest traffic flows into the same rack. Then, the problem becomes how to partition the set of super-VMs for the same job into $K$ parts such that the traffic between each part of the partition is minimized. This problem is equivalent to the well-known minimum $k$-cut problem, which requires finding a set of edges whose removal would partition a graph into $k$ connected components. The VM partition algorithm used here is adopted from the minimum $k$-cut algorithm in \cite{Saran_Vazirani-1995}. For each job $j$, we build a graph $\mathcal{G}_j = (\mathcal{V}_j, \mathcal{E}_j)$, where $\mathcal{V}_j$ represents the set of super-VMs and $\mathcal{E}_j$ represents the traffic between each pair of super-VMs. Then, we compute the Gomory-Hu tree for $\mathcal{G}_j$ and obtain $n_j - 1$ cuts $\{\gamma_i\}$, which contain the minimum weight cuts for all of the super-VM pairs. We remove the smallest $K-1$ cuts from $\{\gamma_i\}$ and obtain $K$ connected components of $\mathcal{G}_j$. For the super-VMs in the same components, we treat them as a super-VM set and assign them into the same rack.

After obtaining all of the partitions of the jobs in every pod, we assign these partitions into racks. For each job, we sort the super-VM sets in decreasing order according to the set size. Subsequently, we assign each set of the super-VMs to racks in a greedy manner. When the assignment of the super-VMs of a job has been completed, we sort all of the racks in increasing order with respect to the number of used servers, and we assign the super-VMs for the next job by repeating the above process, until the super-VMs of all of the jobs have been assigned. Last, we assign the super-VMs for the jobs in set $\mathcal{H}_{N^{pod} + 1}$ to the $N^{pod}$ pods flexibly. Note that this step can be accomplished because $N^{pod}$ is computed by the total resources that are required, and with $N^{pod}$ pods, all of the jobs should be accommodated. The super-VMs of jobs will finally be assigned to the physical servers in each rack.

\begin{figure}[!t]
	\centering
	\includegraphics[scale=0.48]{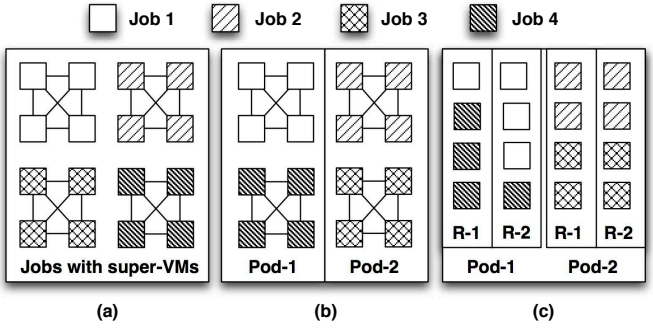}
	\caption{(a) Original jobs' VMs are transformed to super-VMs; (b) the resulting super-VMs are clustered into pods using the $k$-means clustering algorithm; (c) after assigning jobs to pods, the super-VMs are assigned to racks using the minimum $k$-cut algorithm.}
	\label{fig:vma} 
\end{figure}

A simple example that shows the whole process from the moment the job's super-VMs are created until they are assigned to a rack is illustrated in Figure~\ref{fig:vma}. In this example, we have $4$ jobs whose original VMs have been compacted to $4$ super-VMs, as shown in Figure~\ref{fig:vma}(a). Figure~\ref{fig:vma}(b) shows how we cluster them into different pods and, finally, in Figure~\ref{fig:vma}(c), each of the super-VMs is assigned to a rack.

\section{Energy-Efficient Routing}
\label{sec:route}

In this section, we focus on traffic engineering in DCNs to achieve energy conservation. We first explore the relation between energy consumption and routing, and then, based on this relation, we design a two-phase energy-efficient routing algorithm.

\subsection{Exploring Energy-Saving Properties}

As we have discussed in the previous section, in reality we have $R^* > C$. In order to reduce energy consumption, we need to answer the following questions: how many switches will be sufficient and how should the traffic flows be distributed? In this section, we will explore the relation between energy saving and routing, and answer these questions.

The second question can be answered by the following proposition once we have solved the first question.
\begin{proposition}
\label{prop:balance}
With the optimal number of switches determined, the best way to achieve energy savings is to balance the traffic among all of the used switches.
\end{proposition}

This finding is due to the convex manner in which power is consumed with respect to the traffic load. In DCNs, balancing the traffic can be accomplished by many multi-path routing protocols, such as Equal Cost Multi-Path (ECMP) and Valiant Load Balancing (VLB) because data centers usually have networks that have rich connectivity, and these multi-path routing protocols use hash-based or randomized techniques to spread traffic across multiple equal-cost paths. Some more sophisticated techniques, such as Hedera \cite{AL-Fares_Radhakrishnan-2010} and MPTCP (\cite{Han_Shakkottai-2006, Raiciu_Barre-2011}), can also be applied to ensure uniform traffic spread in spite of flow length variations.

To answer the first question, we begin with the aggregation switches (we have shown that nothing can be accomplished with ToR switches once we have the VMs assigned). In general, the following lemma applies.

\begin{lemma}
The optimal energy-efficient routing algorithm will use as few aggregation switches as possible.
\end{lemma}

The same technique can also be applied to the core switches if we ensure that each flow can be routed by the candidate core switches when we choose aggregation switches in each pod. This goal is easy to achieve if we choose aggregation switches from the same positions in different pods and ensure that there will be core switches that connect each pair of them. Taken together, we have
\begin{corollary}
\label{cor:eer}
In the optimal energy-saving solution, the number of active switches is minimized.
\end{corollary}

\subsection{Two-Phase Energy-Efficient Routing}

Based on the answers to the two questions that we asked at the beginning of this section, we devise an energy-efficient routing (EER) algorithm, as presented in Algorithm~\ref{alg:eer}. For each unit of time, we repeat the following two phases. In the first phase, the algorithm aims to find a subset of switches in a bottom-up manner. The estimation of the number of active switches is accomplished by a simple calculation in which we divide the total traffic by the capacity of the switch. However, because it is possible that the multipath routing algorithm might not evenly distribute the traffic flows perfectly, we use the first fit decreasing algorithm, which is a good approximation for the bin-packing problem in which we treat the flows as objects and the maximum transmission rate of the switch as the bin size to ensure that all of the traffic flows can be routed using the selected switches.

\begin{algorithm}[!t]
\caption{\label{alg:eer} \textbf{EER}}
\textbf{Input: } topology $\mathcal{G}=(\mathcal{V},\mathcal{E})$ and VMs assignments\\
\textbf{Output: } routes for flows

\begin{algorithmic}[1]
\FOR{$t \in [t_1,t_r]$}
\STATE Obtain the traffic flows on the network at time $t$ according to the VM assignment
\FOR{$i \in [1,N^{pod}]$}
	\STATE Estimate the number $N_{i}^{agg}$ of the aggregation switches that will be used in the $i$-th pod, and choose them as the first $N_i^{agg}$ switches
\ENDFOR
\STATE Estimate the number $N^{core}$ of core switches that will be used, and choose them
\STATE Use multipath routing to distribute all of the flows evenly on the network formed by the selected switches
\STATE Turn the unused switches into sleep mode
\ENDFOR
\end{algorithmic}
\end{algorithm}

In the second phase, we borrow the most recently proposed multipath routing protocol, MPTCP, to route all of the flows. Compared to the single path routing for each flow in randomized load balancing techniques, MPTCP can establish multiple subflows across different paths between the same pair of endpoints for a single TCP connection. It can be observed that randomized load balancing might not achieve an even distribution of traffic because random selection causes hot-spots, where an unlucky combination of random path selection causes a few links to be overloaded and causes links elsewhere to have little or no load. By linking the congestion control dynamics on multiple subflows, MPTCP can explicitly move traffic away from the more congested paths and place it on the less congested paths. A sophisticated implementation of MPTCP in data centers can be found in \cite{Raiciu_Barre-2011}. The unused switches will be turned into sleep or other power-saving modes in which little power is required to maintain the state.
Because we take advantage of application-level traffic patterns in our model, the network state will remain the same most of the time. Very few state changes will be required, and only on a small number of switches. According to the routes of the flows, the routing tables are generated and sent to corresponding switches at runtime by a centralized controller and an OpenFlow installation in the switches.

\section{Experimental Results}
\label{sec:evaluation}

In this section, we provide a detailed summary of our experimental findings. We associate cost functions to the switches in real data centers, we implement our VM assignment and \textcolor{black}{energy-efficient} routing algorithms presented in the previous sections, and we compare the energy consumption against the solutions obtained by commonly used greedy VM assignment and multi-path routing.

\subsection{Environment and Parameters}

We deploy our framework on a laptop with an Intel Core 2 Duo P8700 $2.53$GHz CPU with two cores and 4 GB DRAM. All of the algorithms are implemented in Python. 

We use two Fat-Tree topologies with $320$ and $720$ switches ($1024$ and $3456$ servers, respectively). The VMs requested by all of the jobs are assumed to be identical, and each server can handle two VMs. For each switch in the data center, a maximum processing speed of $1$ Tbps is given as well as a uniform power function $f(x) = \sigma + \mu x^{\alpha}$ ($x$ is given in Gbps) with $\sigma = 200$ watts, $\mu = 1 \times 10^{-4}$ watts/(Gbps)$^2$ and $\alpha = 2$. Consequently, the maximum power consumption of each switch will be $300$ watts. These parameters define similar commodity switches to the switches discussed at the beginning of Section~\ref{sec:assign}, also meeting the assumption $R^* > C$.

We select a time period of interest $[t_1,t_r]$ such that there are $t_r=100$ minutes with a timeslot length of $1$ minute, during which a set of jobs $\cal{J}$ must be processed in the data center. The set of jobs $\cal{J}$ is generated synthetically, and each job requests a number of VMs that follows a normal distribution $\mathcal{N}(K, 0.5K)$, where $K$ is the number of servers in one rack. Each job is associated with a communication-intensive time interval, which is uniformly distributed during $[t_1,t_r]$. Finally, in each timeslot $t\in[t_1,t_r]$, a traffic matrix $\mathbf{T}_j(t)$ that indicates the traffic between every pair of the VMs of each $j\in\cal{J}$ is provided. The traffic between every pair of VMs follows a normal distribution given by $\mathcal{N}(50 \text{Mbps}, 1 (\text{Mbps})^2)$. The number of jobs is determined by varying the utilization of the servers from approximately $5\%$ to $95\%$ such that VM assignment has significant influence on the energy efficiency of the network and all VMs can be accommodated flexibly. 

\begin{figure}[!t]
	\centering
	\includegraphics[scale=0.5]{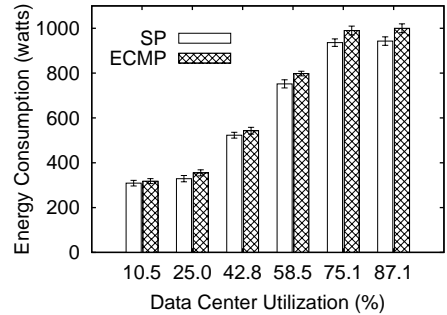}
	\caption{\label{fig:result:ms}Energy consumption under the shortest path routing and ECMP routing algorithms in a Fat-Tree network with $20$ switches. Each value is averaged among $10$ independent tests, and the error bars represent the corresponding standard deviations.}
\end{figure}

\subsection{Benchmarks}

To evaluate the efficiency of our VM assignment algorithm, we compare its results with a greedy VM assignment. This greedy VM assignment algorithm usually assigns an incoming VM to the first server that can serve the computing resources requested by the VM; this method is commonly used in production data centers \cite{Meng_Pappas-2010}.

We also compare our routing algorithm to a shortest-path (SP) routing implemented with Dijkstra's algorithm. We chose SP over ECMP, even though the latter is a multi-path algorithm, because ECMP consumes more power and is far more time consuming. The latter characteristic arises from the need of ECMP to know all of the multiple paths that connect every pair of servers, which could be time-consuming for large-scale topologies and would result in inconvenience when simulating them. With respect to power consumption, ECMP consumes more energy than SP independently of the load on a network. To prove this relationship, both ECMP and SP were run in a small Fat-Tree network (only $20$ switches) several times while varying the amount of load and recording the energy consumption in each case. As shown in Figure~\ref{fig:result:ms}, ECMP always consumed more energy than SP. The extra cost of ECMP came from distributing the load among more paths than SP and using more switches to route the same amount of load, which resulted in higher power consumption.

\subsection{Efficiency of Energy Savings}
\begin{figure}[!t]
	\centering
	\subfigure[]{
 	\label{fig:small:result:16} 
	\includegraphics[scale=0.45]{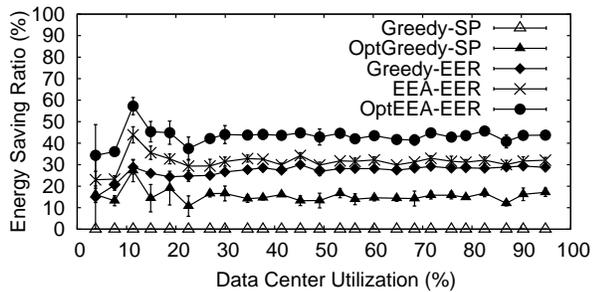}}
	\hspace{-0.1in}
	\subfigure[]{
	\label{fig:small:result:24} 
	\includegraphics[scale=0.45]{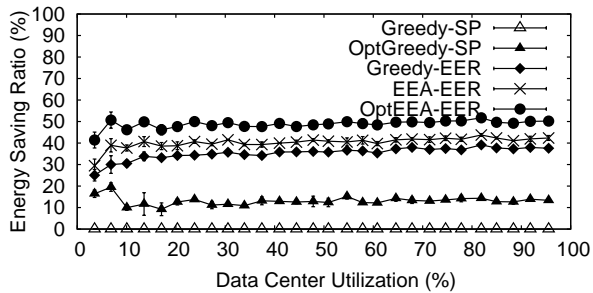}}
	\caption{\label{fig:result}Energy savings ratios under different VM assignment methods and routing algorithms in two data center networks of different sizes with (a) $320$ and (b) $720$ switches. The ratios are obtained as the energy consumption normalized by the amount consumed using Greedy-SP. The values are averaged among $5$ independent tests, and the error bars represent the standard deviations.}
\end{figure}

In this section, we evaluate the performance of the proposed optimized\footnote{The difference between an optimized assignment and a non-optimized assignment is applying or not applying the VM to the super-VM packing transformation.} energy-efficient VM assignment (optEEA) and energy-efficient routing (EER) algorithm by comparing the combination to $4$ different combinations of VM assignment and routing algorithms. Specifically, we compared it with a greedy assignment and SP routing; an optimized greedy (OptGreedy) assignment and SP routing; a greedy assignment and EER; and an energy-efficient VM assignment (EEA) and energy-efficient routing.

These algorithms are tested with two Fat-Tree topologies of different sizes, one with $320$ switches and the other with $720$. For each algorithm and scenario, we vary the load from $5\%$ to $95\%$, and we record the power consumptions to compare the different performances. These results, which are normalized by the Greedy-SP result, are presented in Figures~\ref{fig:result} (a) and (b). From the figures, it can be observed that \\
a) a well-designed VM to super-VM transformation reduces the network energy consumption, as shown in Figure~\ref{fig:result}, by comparing OptEEA-EER with EEA-EER. This arrangement follows the results presented in Proposition~\ref{prop:transformation}. \\
b) EER can save a substantial amount of energy. As seen in the figures, Greedy-EER achieves up to $30\%$ savings compared to Greedy-SP. EER reduces the number of active switches in the network and balances the load among them. Given that the optimal solution, as stated in Proposition~\ref{prop:balance} and Corollary~\ref{cor:eer}, balances the load among a minimum number of active switches, EER can achieve near-optimal solutions. \\
c) Using OptEEA jointly with EER increases the energy savings because they reduce the power consumption in different ways (as explained in Sections \ref{sec:assign} and \ref{sec:route}). It can be seen in Figure~\ref{fig:result} that, regardless of the size of the network, OptEEA-EER outperforms the other algorithms. Combining both algorithms can reduce the energy consumption in the network by up to $50\%$.

\subsection{Running Time}

\begin{figure}[!t]
	\centering
	\subfigure[]{
 	\label{fig:small:time:16} 
	\includegraphics[scale=0.35]{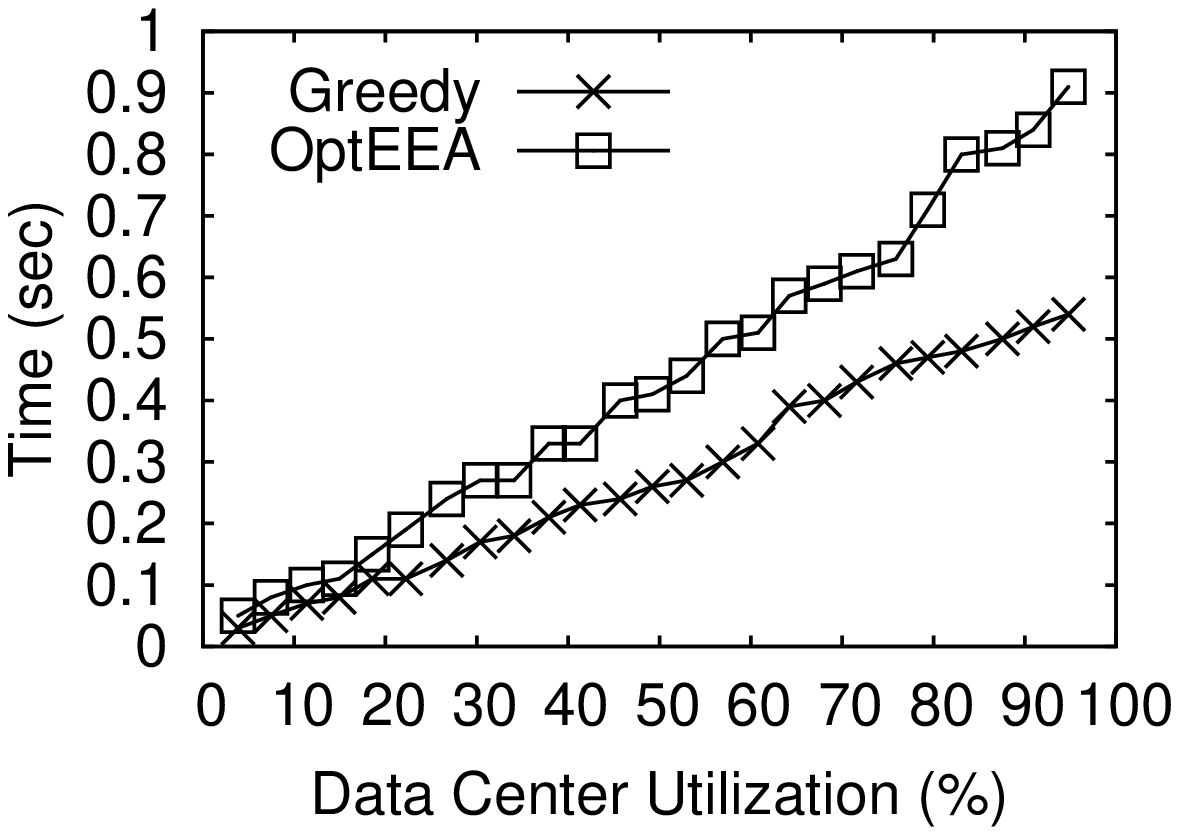}}
	\hspace{-0.3in}
	\subfigure[]{
	\label{fig:small:time:24} 
	\includegraphics[scale=0.35]{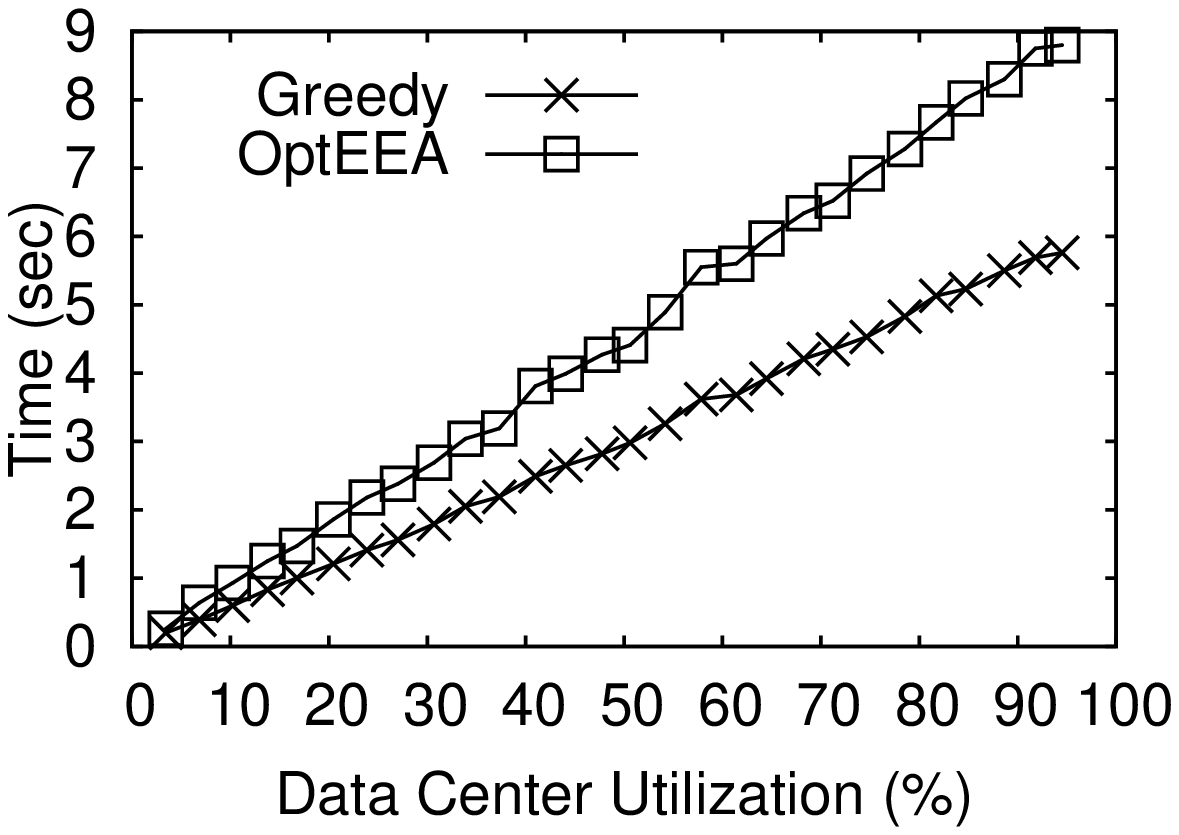}}
	\caption{\label{fig:time}Running times used by the energy-efficient VM assignment algorithm and greedy algorithm in two data center networks of different sizes with (a) $320$ and (b) $720$ switches.}
\end{figure}

We also recorded the running times of the algorithms used in the experiments. These results are presented in Figure~\ref{fig:time}. We find that with the small-scale topology, the proposed algorithm can be completed in one second, while with the large-scale topology, the running time is within 10 seconds. Compared with the greedy algorithm, the running time of the proposed algorithm is only $50\%$ longer. We have also tested the proposed algorithm with large-scale topologies (with tens of thousands of servers connected); most of the time, the running time is bounded by tens of seconds, which is quite acceptable in production data centers.

\section{Discussion}
\label{sec:discussion}

We discuss now some practical problems when applying the whole framework to real data centers.

\textbf{Online Extension.}~~The model and method that we provided in this paper are for offline cases. However, in production data centers, there are most likely cases with dynamic job arrivals or departures. We believe that the proposed algorithms can also be applied to online cases because the jobs are basically assigned to physical machines sequentially. One possible adaption can be that, for each job arrival, we first apply the VM to super-VM transformation, and then we compute the distances between it and the other jobs running in the data center. According to the distances, we assign this job into a pod, and then the remainder of our energy-efficient VM assignment algorithm, as well as energy-efficient routing, can be directly applied. We leave a deliberated adaption to online cases as future work.

\textbf{Implementation.}~~To implement the proposed algorithms in production data centers, we must introduce a centralized job controller to the existing system. This job controller is responsible for determining the VM assignments to the arriving jobs by running optEEA. This job controller is similar to the resource scheduler in current cloud data centers, such as the VMware capacity planner \cite{VCP}. At the same time, it is very easy to envision our VM assignment algorithm being used in a cloud data center by modifying the computing framework (such as Hadoop) to invoke it. However, a centralized network controller is required to run EER. This network controller can be easily adapted from a regular OpenFlow controller because the use of the OpenFlow network appears to be becoming increasingly common in data centers.

\textbf{Impact of Traffic Patterns.}~~Because we want to take advantage of the application-level traffic patterns from upper layers, these patterns could have a strong influence on energy saving efficiency. Recall that the proposed algorithm aims to consolidate VMs with significant traffic. As a result, the more uneven the distribution of traffic between each pair of VMs from a job is, the more energy can be saved. This property is advantageous for MapReduce systems because, when a job is being run, the traffic between different mappers or reducers is negligible, although it is significant between any mapper and reducer. At the same time, optEEA aims to separate jobs that have similar traffic patterns into different pods. Therefore, it is better to have jobs from different applications (different traffic patterns) that run together in the same data center. We leave a comprehensive understanding of the impact of traffic patterns on the efficiency of the whole framework as a future study.

\section{Related Work}
\label{sec:related}

We summarize some related work on network-related optimization problems in data centers, including VM assignment and traffic engineering as well as energy-efficient data center networking.

\subsection{VM Assignment and Traffic Engineering}

Traffic engineering in DCNs has been extensively studied. Because of the centralized environment of data centers, centralized controllers are broadly used to schedule or route traffic flows. Al-Fares \emph{et al.} proposed Hedera \cite{AL-Fares_Radhakrishnan-2010}, which is a scalable, dynamic flow scheduling system that adaptively schedules a multi-stage switching fabric to efficiently utilize aggregate network resources. Benson \emph{et al.} \cite{Benson_Anand-2011} proposed MicroTE, a system that adapts to traffic variations by leveraging the short term and partial predictability of the traffic matrix, to provide fine-grained traffic engineering for data centers. Abu-Libdeh \emph{et al.} \cite{Abu-Libdeh_Costa-2010} realized that providing application-aware routing services is advantageous, and they proposed a symbiotic routing algorithm to achieve specific application-level characteristics.

Recently, data center network virtualization architectures such as SecondNet \cite{Guo_Lu-2010} and Oktopus \cite{Ballani_Costa-2011} have been proposed. Both of them consider the virtual cluster allocation problem, i.e., how to allocate VMs to servers while guaranteeing network bandwidth. In a recent study, Xie \emph{et al.} \cite{Xie_Ding-2012} proposed TIVC, a fine-grained virtual network abstraction that models the time-varying nature of networking requirements of cloud applications, to better utilize networking resources. Meng \emph{et al.} \cite{Meng_Pappas-2010} proposed using traffic-aware VM placement to improve network scalability. Then, they explored how to achieve better resource provisioning using VM multiplexing by exploring the traffic patterns of VMs \cite{Meng_Isci-2010}. In a follow-up study \cite{Wang_Meng-2011}, they investigated how to consolidate VMs with dynamic bandwidth demand by formulating a Stochastic Bin Packing problem and proposed an online packing algorithm. Jiang \emph{et al.} \cite{Jiang_Lan-2012} explored how to combine VM placement and routing for data center traffic engineering and provided an efficient on-line algorithm for their combination. However, they did not consider temporal information on the communication patterns of the applications or the topology features.

\subsection{Energy-Efficient Data Center Networking}

Many approaches have been proposed to improve the energy efficiency of DCNs. These techniques can usually be classified into two categories: The first type of technique is designing new topologies that use fewer network devices while guaranteeing similar performance and connectivity, such as the flatted butterfly proposed by Abts \emph{et al.} \cite{Abts_Marty-2010} or PCube \cite{Huang_Jia-2011}, a server-centric network topology for data centers, which can vary the bandwidth availability according to traffic demands. The second type of technique is finding optimization methods for current DCNs. The most representative work in this category is ElasticTree \cite{Heller_Seetharaman-2010}, which is a network-wide power manager that can dynamically adjust a set of active network elements to satisfy variable data center traffic loads. Shang \emph{et al.} \cite{Shang_Li-2010} considered saving energy from a routing perspective, routing flows with as few network devices as possible. Mahadevan \emph{et al.} \cite{Mahadevan_Banerjee-2011} discussed how to reduce the network operational power in large-scale systems and data centers. \cite{Wang_Zhang-NCA-2013} studied the problem of incorporating rate adaptation into data center networks to achieve energy efficiency. Vasic \emph{et al.} \cite{Vasic_Bhurat-2011} developed a new energy saving scheme that is based on identifying and using energy-critical paths. Recently, Wang \emph{et al.} \cite{Wang_Yao-2012} proposed CARPO, a correlation-aware power optimization algorithm that dynamically consolidates traffic flows onto a small set of links and switches and shuts down unused network devices. Zhang \emph{et al.} \cite{Zhang_Ansari-2012} proposed a hierarchical model to optimize the power in DCNs and proposed some simple heuristics for the model. In \cite{Wang_Zhang-Greenmetrics-2013}, the authors considered integrating VM assignment and traffic engineering to improve the energy efficiency in data center networks. To the best of our knowledge, the present paper is the first paper to address the power efficiency of DCNs from a comprehensive point of view, leveraging an integration of many useful properties that can be utilized in data centers.

\section{Conclusions}
\label{sec:conclusion}

In this paper, we study the problem of achieving energy efficiency in DCNs. Unlike traditional traffic engineering-based solutions, we provide a new general framework in which some unique features of data centers have been used. Based on this framework, we model an energy-saving problem with a time-aware model and prove its NP-hardness. We solve the problem in two steps. First, we conduct a purposeful VM assignment algorithm that provides favorable traffic patterns for energy-efficient routing, based on the three VM assignment principles that we propose. Then, we analyze the relation between the power consumption and routing and propose a two-phase energy-efficient routing algorithm. This algorithm aims to minimize the number of switches that will be used and to balance traffic flows among them. The experimental results show that the proposed framework provides substantial benefits in terms of energy savings. By combining VM assignment and routing, up to $50\%$ of the energy can be saved. Moreover, the proposed algorithms can be run in a reasonable amount of time and can be applied in large-scale data centers.

\section*{Acknowledgments}
The authors would like to thank Dr. Antonio Fern\'andez Anta from Institute IMDEA Networks and other anonymous reviewers for their helpful comments and language editing, which have greatly improved the manuscript. This research was partially supported by NSFC Major International Collaboration Project 61020106002, NSFC\&RGC Joint Project 61161160566, NSFC Creative Research Groups Project 60921002, NSFC Project grant 61202059, the Comunidad de Madrid grant S2009TIC-1692, and Spanish MICINN/MINECO grant TEC2011-29688-C02-01.



\bibliographystyle{IEEEtran}
%

\newpage

\appendix

\noindent\textbf{A. Proof of Theorem 1.}

We prove this theorem by showing that any polynomial-time deterministic algorithm that can obtain the optimal solution for our energy-saving problem can be used to solve QAP. Assume that we are given an instance of QAP with $n$ locations and $n$ facilities. For these locations and facilities, we are also given two matrices $\mathbf{N}_d$ and $\mathbf{N}_c$ of size $n \times n$ to indicate the distance between each pair of locations and the cost between any two facilities, respectively. The total cost of this QAP instance is
\begin{equation}
	\sum_{i_1,i_2 \in [n]} \mathbf{N}_d(i_1,i_2) \mathbf{N}_c\left(\pi(i_1), \pi(i_2)\right), \label{cost_qap}
\end{equation}
where $\pi$ is a permutation of $[n]$. The reduction from QAP to our problem is built as follows: 1) create $n$ nodes for servers; 2) for each pair of servers, connect them with a single switch; 3) for a switch connecting two servers $s_{i_1}$ and $s_{i_2}$, define its power consumption function as $g_{i_1i_2}(x) = \sigma + \mathbf{N}_d(i_1,i_2) x^{\alpha}$, where $\sigma$ is a constant and $\mathbf{N}_d(i_1,i_2)$ is the distance between the $i_1$-th and the $i_2$-th locations in the QAP instance. We treat the facilities as a set of VMs and the $\alpha$ root of the cost between any two facilities $(\mathbf{N}_c)^{1/\alpha}$ as the traffic flow between the corresponding VMs. Therefore, the corresponding energy-saving problem is to allocate each VM into one server such that the total power consumption of the switches is minimized. Given an assignment of VMs $\pi$ (a permutation of $[n]$), the total energy consumption can be expressed by
\begin{equation}
	\sum_{i_1,i_2 \in [n]} \left( \sigma + \mathbf{N}_d(i_1,i_2) \left(\mathbf{N}_c(\pi(i_1), \pi(i_2))^{1/\alpha}\right)^{\alpha} \right). \label{cost_neop}
\end{equation}
It can be verified that the only difference between the total cost of the QAP instance (formula (\ref{cost_qap})) and the total cost in our problem (formula (\ref{cost_neop})) is a constant value $n^2\sigma$. Therefore, when we obtain the optimal solution for our problem, the VM assignment is also optimal for the corresponding QAP. As a result, any polynomial-time deterministic algorithm that optimally solves the energy-saving problem in DCNs can be borrowed to solve QAP. Thus, the proof is complete.
\\\\
\noindent\textbf{B. Proof of Proposition 1.}

Recall that the cost function of each switch is defined as a constant plus the load-dependent part (as shown in Eq.~(\ref{eq:cost_function})). The optimal solution aims to balance the load on the switches because of the convexity of the load-dependent cost.
Hence, the total power consumption of a network is minimized when the power rate of every active switch is minimized. In other words, we choose paths to route flows such that
$f_v(x_v)/x_v$ is minimized for any $v \in \mathcal{V}_a$, where $\mathcal{V}_a \subseteq \mathcal{V}$ is the set of active switches.
This goal can be achieved by choosing all $x_v$ ($v \in \mathcal{V}_a$) to be evenly balanced and as close as possible to $\left(\frac{\sigma}{\mu(\alpha-1)}\right)^{1/\alpha}$, which is denoted as $R^*$.
\\\\
\noindent\textbf{C. Proof of Theorem 2.}

\begin{figure}[!t]
\centering
\includegraphics[scale=0.5]{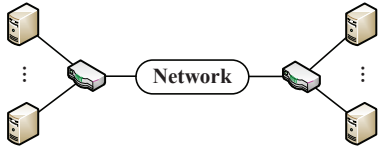}
\caption{\label{fig:tor_reduce_proof} Two ToR switches connected by a network with a general topology.}
\end{figure}

We focus on two arbitrary ToR switches in a Fat-Tree such as the ToR switch in Figure~\ref{fig:tor_reduce_proof}. Let $\mathcal{A}$ and $\mathcal{B}$ represent the set of VMs assigned to the servers connected by the two switches. To conduct our comparison, we assume, without loss of generality, that all of the VMs in set $\mathcal{B}$ can be accommodated into the left-side servers without
exceeding the available resources. Assume that the traffic between each pair of VMs in $\mathcal{A}$ and $\mathcal{B}$ is characterized by a matrix $\mathbf{Q}$, where $\mathbf{Q}(m_1,m_2)$ indicates the traffic flow sent from VM $m_1$ to VM $m_2$. Denote
$$w_1 =  \sum_{m_1 \in \mathcal{A}}\sum_{m_2 \in \mathcal{A}} \mathbf{Q}(m_1,m_2), w_2 = \sum_{m_1 \in \mathcal{A}}\sum_{m_2 \in \mathcal{B}} \mathbf{Q}(m_1,m_2),$$
$$w_3 = \sum_{m_1 \in \mathcal{B}}\sum_{m_2 \in \mathcal{A}} \mathbf{Q}(m_1,m_2), w_4 = \sum_{m_1 \in \mathcal{B}}\sum_{m_2 \in \mathcal{B}} \mathbf{Q}(m_1,m_2).$$

For this configuration, we have in Figure~\ref{fig:tor_reduce_proof}, apart from the power consumed by the other switches, the power consumed by the two ToR switches due to the traffic generated by VMs in $\mathcal{A}$ and $\mathcal{B}$, \textcolor{black}{which} is represented by
$P_1 = 2 \sigma + \mu ( w_1 + w_2 + w_3 )^{\alpha} + \mu (w_2 + w_3 + w_4)^{\alpha}$.
\textcolor{black}{Consider now an alternative assignment in which we move all of the VMs in $\mathcal{B}$ to the left-side servers. On} the one hand, the right-side ToR switch can be shut down to save energy because there is no VM in the right side. Moreover, the power consumption on the intermediate network is reduced because there is no traffic through it. On the other hand, the traffic load carried by the left-side ToR switch increases along with its power consumption. The total power consumed by the two ToR switches is now
$P_2 = \sigma + \mu (w_1 + w_2 + w_3 + w_4)^{\alpha}$.
We now compare the power consumption in both cases. We denote $\Delta P$ as the difference between the two power consumption values. Then, we have

\begin{equation}
\begin{aligned}
\Delta P &\geq P_1 - P_2 \geq \sigma - \mu (w_1 + w_2 + w_3 + w_4)^{\alpha} \\
		&+ \mu (w_1 + w_2 + w_3)^{\alpha} + \mu(w_2 + w_3 + w_4)^{\alpha}.&
\end{aligned}
\end{equation}
Next, we consider the following two cases. \\
\textbf{Case 1:} $\mathbf{\alpha} \geq 2$. Because $\sigma \geq \mu (\alpha - 1) C^{\alpha}$, we have
$\sigma  \geq \mu (\alpha - 1)  C^{\alpha} \geq \mu C^{\alpha} \geq \mu (w_1 + w_2 + w_3 + w_4)^{\alpha}$.
The third inequality follows from $w_1 + w_2 + w_3 + w_4 \leq C$. Then, we have $\Delta P \geq 0$. \\
\textbf{Case 2:} $1 < \mathbf{\alpha} < 2$. We define the function
\begin{equation}
\begin{aligned}
f(w_2, w_3) =&(w_1 + w_2 + w_3)^{\alpha} + (w_2 + w_3 + w_4)^{\alpha}\\
&- (w_1 + w_2 + w_3 + w_4)^{\alpha}.
\end{aligned}
\end{equation}
It is easy to check that the partial derivatives of $f(w_2,w_3)$ are non-negative when both $w_2$ and $w_3$ are non-negative, i.e.,
$\frac{\partial{f(w_2,w_3)}}{\partial w_2} \geq 0$ and $\frac{\partial{f(w_2,w_3)}}{\partial w_3} \geq 0$.
This arrangement means that the function $f(w_2,w_3)$ is monotonically increasing with both $w_2$ and $w_3$. By setting $w_2=w_3=0$, we have
\begin{equation}
\begin{aligned}
	\Delta P	& \geq \sigma + \mu \left(w_1^{\alpha} + w_4^{\alpha} - (w_1 + w_4)^{\alpha}\right) \\
			 & \geq\mu C^{\alpha}\left(\alpha - 1\right) + \mu \left(2\left(\frac{C}{2}\right)^\alpha - C^{\alpha}\right) \\
			  &= \mu C^{\alpha} \left(\alpha + \frac{1}{2^{\alpha - 1}} - 2\right) \geq 0.
\end{aligned}
\end{equation}
The second inequality derives from the fact that $w_1^{\alpha} + w_4^{\alpha} - (w_1 + w_4)^{\alpha}$ is minimized when $w_1 = w_4 = C/2$ with $w_1 + w_4 \leq C$. The last inequality can be easily verified.

In summary, $\Delta P \geq 0$ holds for any $\alpha > 1$, which consolidates the VM results into a more energy-efficient network.
As a result, compacting VMs into racks as tightly as we can and, hence, minimizing the number of ToR switches improves the network energy efficiency. 
\\\\
\noindent\textbf{D. Proof of Theorem 3.}

Given a set of jobs where only one job has significant networking requirements, we focus on the assignment of the VMs for that job. First, we consider an even distribution of these VMs in $k$ different racks. We denote the intra-rack traffic on each ToR switch as $u_i$ $(1 \leq i \leq k)$ and the inter-rack traffic between $2$ racks $i_1$ and $i_2$ as $w_{i_1i_2}$ $(1 \leq i_1, i_2 \leq k)$. We assume that we only use half of the aggregation switches to carry the load evenly.
The total power consumption of the switches in this scenario pod\footnote{The startup cost $\sigma$ is not considered because there are no switches that can be switched off at this point.} is
\begin{equation}
\begin{aligned}
P_1 =& \sum_{i_1=1}^k \mu \left( u_{i_1} + \sum_{i_2 \neq i_1}^k w_{i_1 i_2} \right)^{\alpha} \\
&+ \frac{k}{2} \times \mu \left( \frac{\sum_{i_1=1}^k\sum_{i_2 \neq i_1}^k w_{i_1 i_2}}{2 \times \frac{k}{2}} \right)^{\alpha},
\end{aligned}
\end{equation}
while assigning all of the VMs into a single rack assumes a total power consumption of
\begin{equation}
P_2 = \mu \left( \sum_{i=1}^k u_{i} + \frac{\sum_{i_1=1}^k \sum_{i_2 \neq i_1}^k w_{i_1 i_2}}{2} \right)^{\alpha}. 
\end{equation}
For the sake of tractability, we consider the case in which all of the $u_i$s are roughly equal, with the value denoted as $u$, and all of the $w_{i_1 i_2}$s are roughly identical, with the value denoted as $w$. This scenario is quite common in MapReduce jobs and other cloud computing applications. Define $\Delta P = P_2 - P_1$. Then, we have
\begin{equation}
\begin{aligned}
	\Delta P = 	& \mu \left( ku + \frac{k(k-1)}{2}w \right)^{\alpha} \\
			 &- k \mu \left( u + (k-1)w \right)^{\alpha} - \frac{k}{2} \mu \left( (k-1)w \right)^{\alpha} \\
		\geq	& \mu k^{\alpha} \left( u + \frac{(k-1)w}{2} \right)^{\alpha} \\
			 &- \mu k\left( \left( u + (k-1)w \right)^{\alpha} + \mu \left( (k-1)w \right)^{\alpha} \right) \\
		\geq  & \mu k^{\alpha} \left( u + \frac{(k-1)w}{2} \right)^{\alpha} - k \mu \left( u + 2(k-1)w \right)^{\alpha}\\
		>  & (k^{\alpha} - k4^{\alpha})  \mu \left( u + \frac{(k-1)w}{2} \right)^{\alpha}
		>   0,
\end{aligned}
\end{equation}
where the second inequality is due to the convexity of the power consumption incurred by traffic loads, and the last inequality derives from our assumption that $k \geq 4^{\frac{\alpha}{\alpha - 1}}$. Thus, as long as $K \geq 4^{\frac{\alpha}{\alpha - 1}}$, it is possible to distribute all of the VMs \textcolor{black}{among the} servers in one pod to reduce the power consumption.
\\\\
\noindent\textbf{E. Proof of Theorem 4.}

Assume that we have one job with all of its VMs assigned to a single pod $A$. Next, consider moving some VMs from $A$ to a new pod $B$. Consider the simple case in which we move the VMs from a whole rack in $A$ to an empty rack in $B$. Fat-Tree has the property that, for each aggregation switch, the outer fan-out (to other pods) is not larger than the inner fan-out (to ToR switches within the pod). Then, the number of node-disjoint paths between any two ToR switches in $A$ will not be smaller than the paths between two ToR switches in $A$ and $B$. As a result, moving the VMs from a whole rack to a different pod will never reduce the traffic on any ToR or aggregation switch; it will bring extra traffic in core switches, without reducing the power consumption.
\\\\
\noindent\textbf{F. Proof of Proposition 2.}

The proof can easily be seen from the fact that the traffic between any VMs assigned to the same server does not go to the physical NICs on the host server as well as the network. Then, it is quite natural to assign VMs for the same job to the same servers to reduce the network traffic. In this sense, compacting VMs with high communication traffic will reduce the traffic on the network, resulting in more energy savings.
\\\\
\noindent\textbf{G. Proof of Lemma 1.}

We focus on the aggregation switches in one pod. Recall that in the Fat-Tree topology, the connectivity between ToR switches and aggregation switches is supported by all-to-all mapping links. Thus, we can choose any aggregation switch to carry any ingress or egress flows of a ToR switch. We denote the minimum number of aggregation switches to be used as $N^{agg}$. We will show that for any $n \geq N^{agg}$, the minimum total power consumption of the aggregation switches obtained using $n$ aggregation switches will always be smaller than that obtained by using $n+1$ aggregation switches.

Assume that in the optimal solution with $n$ aggregation switches, the total load going through the $i$-th aggregation switch is $p_i \in (0, C]$ $(1 \leq i \leq n)$, while using $n+1$ aggregation switches, this value is $q_i \in (0,C]$ $(1 \leq i \leq n +1)$. Because all of the switches are identical, without loss of generality, we assume that $q_{n+1}$ is the highest load among all $q_i$ and that $p_i$ and the remaining $q_i$ are sorted in descending order. Denoting $\delta_i = p_i - q_i$ for $1 \leq i \leq n$, we have $q_{n+1} = \sum_{i=1}^n \delta_i$. Because both solutions are optimal, it can be observed that $\delta_i \geq 0$ for $1\leq i \leq n$. Using $n$ switches, the total power consumption is presented as
\begin{equation}
P(n) = n \sigma + \mu \sum_{i=1}^n p_i^{\alpha},
\end{equation}
While using $n+1$ switches, the total power consumption is
\begin{equation}
P(n+1) = (n+1) \sigma + \mu \sum_{i=1}^n (p_i - \delta_i)^{\alpha} + \left( \sum_{i=1}^n \delta_i \right)^{\alpha}.
\end{equation}
To complete the proof, it is sufficient to show that for any $n \geq N^{agg}$, we have $P(n+1) \geq P(n)$. Denoting the difference between the two optimal solutions as $\Delta P$, then
\begin{equation}
\begin{aligned}
	\Delta P	= & P(n+1) - P(n) \\
			= & b + \mu \sum_{i=1}^n ((p_i-\delta_i)^{\alpha} - p_i^{\alpha}) + \mu \left(\sum_{i=1}^n \delta_i\right)^{\alpha}.
\end{aligned}
\end{equation}
Note that $\Delta P$ is a function of the variables $\vec{p}$ and $\vec{\delta}$, where $\vec{p} = (p_1, p_2, ...p_n)$ and $\vec{\delta} = (\delta_1,\delta_2,...,\delta_n)$. Because $\vec{p}$ and $\vec{\delta}$ are independent and $\delta_i \geq 0$, $\Delta P$ is minimized when we set $\vec{p} = (C,C,...,C)$. In other words,
\begin{equation}
\begin{aligned}
\Delta P	\geq & \sigma + \mu \sum_{i=1}^n ((C-\delta_i)^{\alpha} - C^{\alpha}) + \mu \left( \sum_{i=1}^n \delta_i \right)^{\alpha} \\
		\geq& \sigma + \mu \left( \sum_{i=1}^n \left( \frac{nC}{n+1} \right)^{\alpha} - \sum_{i=1}^n C^{\alpha} \right) \\
		\geq& \mu C^{\alpha} + \mu \left( (n+1)\left( \frac{nC}{n+1} \right)^{\alpha} - nC^{\alpha} \right) \\
		=& \mu C^{\alpha} \left( (\alpha - 1) + n \left( \left( \frac{n}{n+1} \right)^{\alpha} - 1   \right)  \right) \\
		\geq &\mu C^{\alpha} \left( \alpha - 1 + n\left( \frac{n+1}{n+\alpha} - 1 \right) \right) \\
		=& \mu C^{\alpha} \frac{\alpha(\alpha - 1)}{n + \alpha} > 0,
\end{aligned}
\end{equation}
when $\alpha > 1$. The second inequality derives from the fact that $\Delta P$ is minimized when we set $\vec{\delta} = \left( \frac{C}{n+1}, \frac{C}{n+1},..., \frac{C}{n+1} \right)$. The third inequality is due to the restriction that $\sigma \geq \mu C^{\alpha}(\alpha - 1) > 0$. The fourth inequality is obtained by applying the necessary condition on $n$ that the first derivative equals zero. Having $\Delta P> 0$ means that using fewer aggregate switches results in better energy efficiency. 

\nobalance

\end{document}